# Tracking areas with increased likelihood of surface particle aggregation in the Gulf of Finland: A first look at persistent Lagrangian Coherent Structures (LCS).

Andrea Giudici [a,*], Kabir A. Suara [b], Tarmo Soomere [a,c], Richard Brown [b]

[a] Department of Cybernetics, School of Science, Tallinn University of Technology, Akadeemia tee 21, 12618 Tallinn, Estonia

[b] Environmental Fluid Mechanics Group, Queensland University of Technology (QUT), QLD, 4000, Australia

[c] Estonian Academy of Sciences, Kohtu 6, 10130 Tallinn, Estonia

[*] Corresponding author, e-mail: andrea.giudici@ttu.ee

**Abstract.**

We explore the possibility to identify areas of intense patch formation from floating items due to systematic convergence of surface velocity fields by means of a visual comparison of Lagrangian Coherent Structures (LCS) and estimates of areas prone to patch formation using the concept of Finite-Time Compressibility (FTC, a generalisation of the notion of time series of divergence). The LCSs are evaluated using the Finite Time Lyapunov Exponent (FTLE) method. The test area is the Gulf of Finland in the Baltic Sea. A basin-wide spatial average of backward FTLE is calculated for the GoF for the first time. This measure of the mixing strength displays a clear seasonal pattern. The evaluated backward FTLE features are linked with potential patch formation regions with high FTC levels. It is shown that areas hosting frequent upwelling or downwelling have consistently stronger than average mixing intensity. The combination of both methods, FTC and LCS, has the potential of being a powerful tool to identify the formation of patches of pollution at the sea surface

**Keyword**s: patchiness, marine pollution, Lagrangian coherent structures, flow compressibility, Finite-Time compressibility, surface pollution

**Research highlights**

- Lagrangian Coherent Structures are calculated in the Gulf of Finland (GoF)
- A basin-wide spatial average of backward FTLE is calculated for the GoF
- Mixing strength displays a seasonal pattern

- Backward FTLE features loosely resemble those of high-Finite Time Compressibility regions.
- Mixing in the hotspot areas of up/downwelling is consistently higher than the average

**1. Introduction**

The Gulf of Finland (GoF) is an actively investigated and heavily exploited area of the Baltic Sea that is a particularly sensitive brackish water body under extremely strong anthropogenic pressure (Leppäranta and Myrberg, 2009). The GoF hosts a plethora of interesting phenomena that range from small-scale vortices, fronts and intense vertical motions up to large-scale circulation (Soomere et al., 2008). The presence and complex interactions between different drivers, and across all seasons, result in a variety of curiosities, from narrow frontal areas with extremely high bioproduction (Pavelson et al., 1997) to upwelling-driven extensive toxic cyanobacterial blooms (Wasmund et al., 2012) and occasional export of saltier deep waters (Elken et al., 2014). This water body exhibits a complicated hydrography – typical of estuaries –, with saline input from the Baltic Sea proper on the West side and a large freshwater input from rivers on the East side. The transient nature and high variability of the patterns of its driving forces, extreme complexity of the dynamics of the marine environment, very small internal Rossby radius (usually 1–3 km, Alenius et al., 2003), chaotic appearance and low directional persistence of surface currents (Andrejev et al., 2004) leads to strong horizontal and vertical variations in the hydrodynamic regime of this water body (Maljutenko and Raudsepp, 2019) that is expressed *inter alia* via extremely complicated paths of drifters (Delpeche-Ellmann et al., 2017).

Other ageostrophic ocean processes such as frontogenesis, submesoscale mixed-layer instabilities, shelf break fronts and topographic interactions on the continental shelf are known to produce surface-divergent flows, which affect buoyant material over time (Jacobs et al., 2016).

In particular, up- and down-welling phenomena often take place in the area (Hela, 1976; Hela and Haapala, 1994; Lehmann et al., 2012) mainly owing to frequent relatively strong alongshore winds. Upwelling phenomena play an important role in bringing cold water from deeper layers to the surface, in mixing water masses and in generating frontal areas (e.g., Lips et al., 2009). Downwelling takes place at the opposite coast, to that where upwelling takes place, creating a circulation across sections of the gulf, whenever alongshore winds blow steadily for extended periods of time (greater than 60 h) (Myrberg et al., 2010). Such phenomena are typically associated with steep sea surface temperature (SST) variations, as well as with patchiness (sometimes called also *patchness*) (Brentnall et al., 2003; Kononen et al., 1992; Granskog et al., 2005) of different quantities at the surface. The latter is generally referred to in the literature as a localised increase or decrease in the concentration of any biotic or abiotic fields from that of the ambient level. The magnitude of variations in the relevant concentration field may be very different for different properties of the sea. For example, changes in the temperature by a fraction of degree or in salinity by a fraction of per million often deserve attention whereas cyanobacteria blooms seem fairly homogeneous even if the amount of cyanobacteria per unit of sea surface varies several times. For this reason, we specify the threshold for detection of patches below based on the actual variations of evaluated quantities.

These facts, together with very dense marine traffic, among the most intense in the world (Soomere et al., 2014), makes the gulf a very interesting test bed to study the dynamics of floating litter. A natural driver of the generation of patches of litter (understood here as areas with increased concentration of litter) is a system of three-dimensional (3D) motions of the sea, acting upon a two-dimensional (2D) surface of particles floating at the surface. See, e.g., Giudici et al. (2019) for an overview of the related effects. This kind of behaviour is naturally triggered when some matter is concentrated in the surface layer (Lee, 2010; Froyland et al., 2014) or motile species that inhabit this layer (Chatterjee et al., 2016) in marine environment. In addition, at the micro-scale level, active buoyancy regulation commonly leads to intense patchiness for non-motile species as compared to passive tracers (Arrieta et al., 2017): indeed the local convergence or divergence of the surface velocity field may lead to the generation of contracting (e.g., during a downwelling) or expanding (during an upwelling) mechanisms among the floating items (Fennel et al., 2010; Giudici et al., 2012; Hernandez-Carrasco et al., 2018).

These areas hosting contracting/expanding movements often seem to appear in unstructured patterns, and major patchiness phenomena are unlikely to take place unless they are enhanced by biological chains (Messie and Chavez, 2017) or specific features that are sometimes active in the periphery of intense vortices (Samuelsen et al., 2012; Zhurbas et al., 2019).

Diagnosing regions likely to contain clusters, without the need to integrate millions of particle trajectories, is not a trivial task. A convenient solution is to follow material deformation lines. Material deformation has previously been decomposed into dilation and area-preserving stretch processes in order to reduce the computational footprint (Huntley et al., 2015).

By means of a generalization of the definition of divergence (Kalda et al., 2014), the likelihood of the generation of patches at the (sea) surface has been characterised to some extent using the concept of flow compressibility of the velocity field (Giudici et al., 2012). This measure is not related to the compressibility of water masses. Instead, it is employed as a generalisation of the notion of divergence of velocity field and used as a convenient (albeit local in time and space) measure of the "ability" of certain marine areas to serve as a gathering place of floating items or locations with increased concentrations of substances that are forced to remain at the sea surface (Pichel et al., 2007).

The classical notion of compressibility of a velocity field relies on the (Helmholtz) decomposition of realistic fields of ocean currents into their solenoidal (with zero divergence) and potential (with zero curl) components. Flow compressibility is the relative weight of the potential component of the velocity field.

This decomposition and definition rely on instantaneous properties of the flow field and ignore its history. In reality, the history of the (interrelations) of the divergence field and Lagrangian transport of the parcels of a medium do play a large role as demonstrated, for example, by Samuelsson et al. (2012). To account for the interplay of these two drivers, a slightly different definition of (surface) flow compressibility called finite-time compressibility and denoted as FTC was introduced in (Kalda et al., 2014). This notion to some extent takes into account the history of processes on the sea surface (Kalda et al., 2014). It has therefore potential to reveal the location and the temporal persistence of areas, which regularly exhibit high chances of increases in the surface concentrations of various items and substances. This measure generalises the analysis of (Samuelsen et al., 2012) towards systematically to quantifying the joint impact of convergence fields moving along the sea surface, and the Lagrangian transport of the growing patch in the surface layer on the concentration of floating matter resembles that.

The compressibility of the surface velocity field is evaluated semi-locally, by observing the relative changes in surface elements that are passively carried by surface currents during a certain time interval. Specifically, the local instantaneous values of FTC are evaluated using the ratio of the relative changes of the surface area of a triangular floating element (made of a triplet of simulated tracers) over the similar changes to their edges (Giudici et al., 2012). The relative change $dS_t$ of the area $S_t$ of such element over one time step is $dS_t = (S_t - S_{t-1})/S_t$. The relative changes to the squared lengths of two of its edges $A_t$ and $B_t$ are $dA_t = (A_t^2 - A_{t-1}^2)/A_t^2$ and $dB_t = (B_t^2 - B_{t-1}^2)/B_t^2$. The approximate values of FTC over a certain time interval are calculated as follows (Giudici et al., 2012):

$$C_{FTC} = \frac{2dS_{rms}}{dA_{rms} + dB_{rms}}, \tag{1}$$

where $dS_{rms}$, $dA_{rms}$ and $dB_{rms}$ stand for the root mean square of the instantaneous values of $dS_t$, $dA_t$ and $dB_t$ over the time interval during which the motion of the element is tracked. The tracking time for each estimate is chosen so that the result characterises a fixed small region of the sea surface that contains three grid cells of the relevant ocean model. These local instantaneous values are further averaged over a large number of realisations of the flow that start at different time instants and used to evaluate the FTC field of the entire target area.

The procedure of calculations of FTC in (Giudici et al., 2012; Kalda et al., 2014) was as follows. The chosen time interval $t_D$ of interest (for example, 1 month) is divided into shorter (and optionally overlapping) time spans (windows). These windows are shifted by at least 12 hrs to remove serial correlations between subsequent simulations. The average time of the drift of floating substances to the nearshore is about 5.3 days in the GoF (Andrejev et al., 2011). This means that each time span should much shorter than this time scale (e.g., ≤3 days) to avoid distortions to the triangular elements by the presence of the shores. Using this technology, nine areas of the GoF were identified, which stably show increased values of FTC compared to the background level (Giudici and Soomere, 2013). These areas might owe their existence to the turbulent interactions between several drivers (Messie and Chavez, 2017; Vucelja et al., 2007; Balk et al., 2004; Denissenko et al., 2006; Martin, 2003; Delpeche-Ellmann et al., 2017). The impact of wind together with the ability of pieces of litter to stick together may accelerate the growth of patchiness. It was shown that the interaction between currents and winds does lead to an increase of the patchiness of the areas. The size of the largest patches is governed universally (Giudici, Kalda, Soomere, 2019) by a stretched-exponential power law $f(x) = A\exp(-x^b)$.

The dispersion and aggregation of floating litter on sea surface is usually governed by the combination of local weather events and the interactions of the flow fields with local morphological features, ranging across a wide temporal scale. The overlap of several processes, operating at several different scales, existing in such system can lead into the chaotic nature of material transport at the sea surface (Suara et al., 2017, 2019).

As with any turbulent flow, material coherence emerges ubiquitously in fluid flows. It admits distinct signatures in virtually any diagnostic scalar field or reduced-order model associated with these flows. In the conditions of the GoF such signatures may exhibit anomalous (for the northern hemisphere) anticyclonic patterns in terms of Lagrangian transport (Soomere et al., 2011; Maljutenko and Raudsepp, 2019) These signatures, however, do not reveal the root cause of flow coherence. The approach of Lagrangian Coherent Structures (LCS) seeks to isolate the root cause by uncovering special surfaces of fluid trajectories that organise the rest of the flow into ordered patterns (Olascoaga et al., 2006; Haller, 2015), and can be used as an effective tool to forecast sudden changes in environmental pollution patterns (Olascoaga and Haller, 2012). LCS renders lines (surfaces), occurring in chaotic two (three) dimensional flows, which serve as distinctive barriers that cannot be crossed by an ideal tracer (Haller, 2001). They are the local strongest repelling or attracting material lines (surfaces).

The trajectories of floating particles are generally sensitive to their initial conditions, making the assessment of flow models (as well as observations of single tracer trajectories) quite unreliable (Vandenbulcke et al., 2009). Behind such complex and sensitive tracer patterns in the study area that is frequently affected by energetic upwelling flows that modify even the Ekman transport (Delpeche-Ellmann et al., 2017), however, LCS are expected to represent a robust foundation of material surfaces, which play a key role, in shaping those trajectories, or even greatly affecting the sea dynamics of an area (Olascaoga, 2010).

Albeit the description and interpretation of turbulent mixing mechanisms is not an easy task, the identification of such material lines, which act as barriers for repulsion or attraction for clusters of floating particles, can be helpful in the further identification of areas of accumulation of pollutants. Uncovering such structures from measured and modelled flow data promises a simplified way to understand the overall flow geometry, a better quantification of the material transport, as well as an opportunity to forecast, or even influence, large-scale flow features and mixing events.

A lot of work has been done with respect to techniques used to identify LCS (Hadjighasem et al., 2017), and their application to velocity fields from hydrodynamic models and is still an open area of research. In this work, we apply the finite-time Lyapunov exponent (FTLE) as a proxy to identify potential LCS. We make a first attempt at characterising them, with respect to their appearance pattern and seasonality within the Gulf of Finland. Specifically, the work presented here is aimed at verifying the seasonality in the dynamics of areas prone to patch formation as well as the presence and characteristics of hotspots for upwelling and downwelling events in the this water body.

**2. Data and Methods**

Initial studies on patch generation in the area (Giudici et al., 2012; Giudici and Soomere, 2013; Kalda et al., 2014) were based on the output of a medium-resolution ocean circulation model designed for climate studies (Meier et al., 2003; Meier, 2007), which was only to some extent able to properly resolve eddies and therefore only partially able to represent the dynamics of the area correctly. The velocity fields used in subsequent studies and this paper are obtained by means of the OAAS (Andrejev and Sokolov, 1989; 1990) hydrodynamic model, as carried out in the framework of BONUS+ BalticWay cooperation (Soomere et al., 2014), with a spatial resolution of 1 nautical mile (NM). In order to have our results compatible with the research into intense patch-formation areas and properties of the emerging patches (Giudici and Soomere, 2013, 2014; Giudici et al., 2018, 2019), we employ the same time period of surface velocity fields for the GoF.

The OAAS hydrodynamic model is a primitive-equation, time-dependent, free-surface, baroclinic model based on hydrostatic approximation. It was specifically designed for simulating the circulation in basins with complex geometry. The vertical grid resolution is 1 m (excluding the uppermost layer which is 2 m thick). The model area was that of the inner Gulf of Finland, to the east of longitude 23°27' E. A detailed description of the main features of the model is presented in (Andrejev et al., 2011). The model run was forced with the river runoff data from Bergström and Carlsson (1994), and meteorological data from a regionalization of the ERA-40 re-analysis over Europe. This data is based on a regional atmosphere model with a spatial horizontal resolution of 25 km, which makes use of adjusted wind information to match measured statistics (Höglund et al., 2009; Samuelsson et al., 2011). The employed data covers the time span of 1989–1992, with a temporal data resolution of 3 h (even though internally, calculations are performed with a resolution of 60 s).

The ridge of high finite-time Lyapunov exponent (FTLE) values, which visualises potential LCSs and identifies parts of the flow field that accumulate/spread materials more than others, is applied. More robust and exact methods for identification of LCS are in the literature (Haller, 2011). Locating such structures would normally require a detailed assessment of the stability of material surfaces in the flow domain of interest. A first-order approximation to this problem is to detect those material surfaces, along which infinitesimal deformation is locally maximal or minimal. FTLE fields are computed from the surface layer of the available discrete velocity data set from the hydrodynamic model described above.

Flow map is obtained by advecting tracer particles for a finite time between initial time $t_0$ and final time $t_1$, using a Runge-Kutta (4,5) scheme. We use a Matlab realisation of this scheme – a 4th order time-adaptive solver, which dynamically optimises the time step to meet an absolute integration tolerance of $10^{-6}$. The FTLE field is computed as $FTLE(\vec{x}, t, \tau) = \frac{1}{|\tau|} \ln \sqrt{\lambda_{max}(\Delta)}$, where $\tau = t_1 - t_0$ is the advection time and $\lambda_{max}$ is the largest eigenvalue of the Cauchy-Green deformation tensor $\Delta$ obtained from the flow map. The computation was performed using scripts modified from BarrierTool (Katsanoulis and Haller, 2019). The line of the FTLE field which is the curve that goes through the local maxima of the FTLE evaluated along the direction of fastest change in the FTLE values[1] and signifies the potential LCSs (Haller, 2001). This line is readily identified from visual examination and changing the threshold values of the FTLE contour plots (Rockwood et al., 2019). Particles are advected backward ($\tau > 0$) in time to reveal the attracting/unstable material lines. Attracting material lines represent lines of maximum accumulation, so particles that are initially further apart come together rapidly. A free-slip condition is applied at the land-water interface, therefore particles that cross the land boundaries are removed from the calculation. The backward FTLE trajectory integration allows to identify attractor in the flow, therefore renders the areas of accumulation.

---

[1] Lines of curvature in terms of differential geometry: lines corresponding to the minimum (maximum) normal curvature of the surface and going through a (local) maximum (minimum) of the surface.

To identify potential areas of strong accumulation in the Gulf of Finland, the calculated FTLE fields are normalised by the maximum FTLE values. Similarly, analysis of persistence of accumulation within the basin is performed considering points with normalised FTLE above a threshold of 0.71 for identifying the FTLE ridge. Though the thresholding results of FTLE narrows the ridges of FTLE, it does not change their location (Rockwood et al., 2019). Normalised FTLE threshold as low as 0.5 has been used in literature to identified ridges of FTLE (Dauch et al., 2018).

## 3. Results

Initially, we calculated maps of backward FTLE covering the entire Gulf of Finland,. These maps (Fig. 1) are a simple and efficient tool to visualise hyperbolic LCS amid a plethora of physical complexities. The dynamical picture rendered by FTLE field is dependent on the integration time, and this should be selected to reflect the process of interest (Huhn et al., 2012; Suara et al., 2020). Herein, an integration time of three days is selected. This time interval is longer than the diurnal wind pattern across the gulf (Soomere and Keevallik, 2003; Savijärvi et al., 2005; Soomere et al., 2008) and shorter than the time for most of the tracers to reach any coastal segment as described above (Viikmäe et al., 2010; Soomere et al., 2014).

The resulting maps are intriguing snapshots of the turbulent features of the Gulf of Finland when coupled with a sliding time window. A movie of the FTLE field over a year period is provided in the Supplementary material (SV1). The visualisation in the video as well as Fig. 1, in which the FTLE fields at some instances of windy and calm periods are compared, suggests a seasonal dynamics of the gulf which is influenced by the wind. This feature apparently is the main reason behind seasonally different optimum fairways in terms of minimum environmental effects of possible accidents on the nearshore via current-driven transport of pollution (Murawski and Woge Nielsen, 2013; Soomere et al., 2014).

To further examine the patterns in the variation of the dynamics in the basin, we characterise the overall clustering potential of the GoF, by employing a spatial average of such FTLE features. Because an increase in eddy activity implies a simultaneous increase in both dispersion and attraction of material therefore, only the *backward* FTLE field $FTLE^-$ which signals accumulation is analysed here (Suara et al., 2020). This average would not smear out the mixing effect but rather is intended to quantify the overall clustering effect, relative to other time instants. Following d'Ovidio et al. (2004), the temporal variation of such spatially averaged positive FTLE features quantifies the overall mixing strength in a system. Figure 2 shows a plot of such a measure over a time range spanning from 1989 to 1992 and the instantaneous and average wind speed during this time period. This plot signals that the clustering potential of certain offshore areas of the GoF steadily increases, and peaks, during the windy season, and then decreases, during the calm season.

Figure 3 quantifies the cross-correlation of the wind intensity, and the mean FTLE signal. While the correlation coefficient for their instantaneous values is about 0.6, it reaches a maximum of $r = 0.91$ when the two signals are offset by around 20 h. This feature apparently reflects an offset between strong wind events and the formation of clustering areas in the natural environment. It is likely that the time scale of this offset is site-specific and reflects properties of local dynamics of water masses.

Under these conditions, the minima of the clustering potential are all reached during calm time periods. This course matches well the perception that surface clustering in the GoF is strongly augmented by the combined effect of currents and winds (Giudici et al., 2018).

From this metric produced for the entire Gulf of Finland, and over such a time range, we made an attempt to find areas in the basin, which exhibit consistently high FTLE values, across an entire calendar year. The results were compared with the set of regions, identified in (Giudici and Soomere, 2013), which are characterised by high-FTC values during the same time frame (Figure 4).

The locations of high values of backward and forward FTLE signify the areas for strongest clustering and spreading potentials, respectively. Because of the difference in the dynamics between consecutive temporal windows, the FTLE fields are normalised by the basin-wide maximum FTLE value as recommended in (Peikert et al., 2014). Therefore, a value close to one signifies areas of strong activity of clustering or spreading. For a year, FTLE calculations performed using 12-h interval between consecutive windows were used, resulting in 718 maps. For each FTLE map, locations with the normalised FTLE values greater than the threshold value, are assigned a value of 0.5 day (0 for otherwise) and cumulated over the whole year. Rockwood et al. (2019) utilised threshold values >0.55 to identify FTLE ridges and showed that the selection of the threshold does not alter the locations but narrow the ridges of the FTLE. Consistently, the threshold values used in this paper did not alter the entire patter but possibly narrowed the areas of strong activity identified with high frequency of occurrence of high FTLE values.

Following (Shadden, 2011), we set the threshold for "high" FTLE values in this normalisation to 0.7. The number of days per year displaying values of normalised FTLE above this threshold of 0.7 is shown on Figure 5. The average number of days per year that display above-threshold FTLE values on wet grid cells is about 2 days. This count has very large spatial variations and reached its maximum values of about 150 days per year for the time period considered. Some of the visually distinguishable areas of systematically large values of this count match well the high-FTC areas (areas 2, 3, 4, 6, and 7 in Figure 4). However, other regions of high FTC (areas 1, 5 and 8 in Figure 4) do not exhibit consistently high FTLE values. This feature signals that the presence of high values of FTC does not necessarily lead to the attraction of floating items to the relevant area. It also appears to support the view that, at least some of these locations are, irrespectively of the season, active hot spots for clustering phenomena in the GoF.

### 4. Discussion

In this paper, Lagrangian coherent structures (LCSs) for the micro-tidal Gulf of Finland were calculated for the first time. The dynamics of water masses in this water body is strongly time dependent, largely driven by air pressure and wind, and often exhibits almost chaotic patterns (Soomere et al., 2008). In essence, we have made an attempt to evaluate the applicability of the concept of LCS for this water body using a Finite Time Lyapunov Exponent (FTLE) diagnostic approach to describe the accumulation areas of floating material in the system. The transport boundaries of material accumulation identified through LCS in the context of this work are not necessarily 'perfect'. Rather they indicated regions within the domain with higher accumulation properties relative to others.

While gulf-wide snapshots of LCS were found to be highly dynamic, a spatial average of the positive (attracting) FTLE features, calculated for the entire basin, shows a discernible seasonal course in the corresponding overall mixing strength. The seasonal pattern is visible over the considered time span 1989–1992. The pattern is stronger during the windy season and somewhat weaker during the calm season. The instantaneous correlation coefficient between the mean FTLE signal and the wind is about 0.6, This correlation is much higher ($r = 0.91$) between the FTLE and wind properties that occurred about 20 hearlier. This feature apparently characterises the reaction time of Lagrangian transport to wind properties in the study area.

A partial match (5 out of 8 areas) was found in visually comparing the previously investigated areas of high levels of finite-time compressibility (FTC, prone to accumulation of floating items) in the Gulf of Finland, and regions where the newly calculated mean FTLE values were consistently above threshold, during the years 1987–1991. This partial match further supports the view that the areas identified using the concept of finite-time compressibility (Giudici and Soomere, 2013; 2014) display increased clustering potential and represent probably up/down-welling hotspots.

On the one hand, the overall coherence of the results shown with the ones obtained using the concept of FTC suggests that both methods, FTC and LCS, have the potential of being a powerful tool to provide useful insights into the long-term effects of turbulent flows which govern the formation of patches of pollution at the sea surface. On the other hand, there are several fairly large differences between the results of the two methods. In particular, the majority of high-FTLE locations are in the north of the gulf where no high-FTC regions were identified in the earlier studies. This mismatch may stem from the differences in the procedure of specification of the high-FTC regions where the areas that showed high FTC values only during single seasons were excluded. As the high-FTC areas are commonly associated with downwelling events, such areas may have also strong interannual variability depending on the relative frequency of westerly winds (that drive downwelling in the southern part of the gulf) and easterly winds (that support downwelling in the northern part).

There also exist some limitations to the undertaken approach. Most notably, without some more advanced criteria (as shown for example in Haller, 2002, 2011; Tang et al., 2011; Farazman and Haller, 2012), the employed FTLE approach to calculate LCS is heuristic. It ignores the direction of largest stretching, which may be along or close to directions tangent to the initial trajectory. In addition to that, FTLE ridges mark hyperbolic LCS positions which are used to extrapolate information on the mixing strength, but also highlight surfaces of high shear (Haller, 2002).

Although FTLE plots are popular visual diagnostic tools of Lagrangian coherence, more reliable mathematical methods have also been developed for the explicit identification of LCSs, as parameterised material surfaces. Such methods should be taken into consideration in further investigations of appearance patterns of LCS in the Gulf of Finland.

**List of Figures**

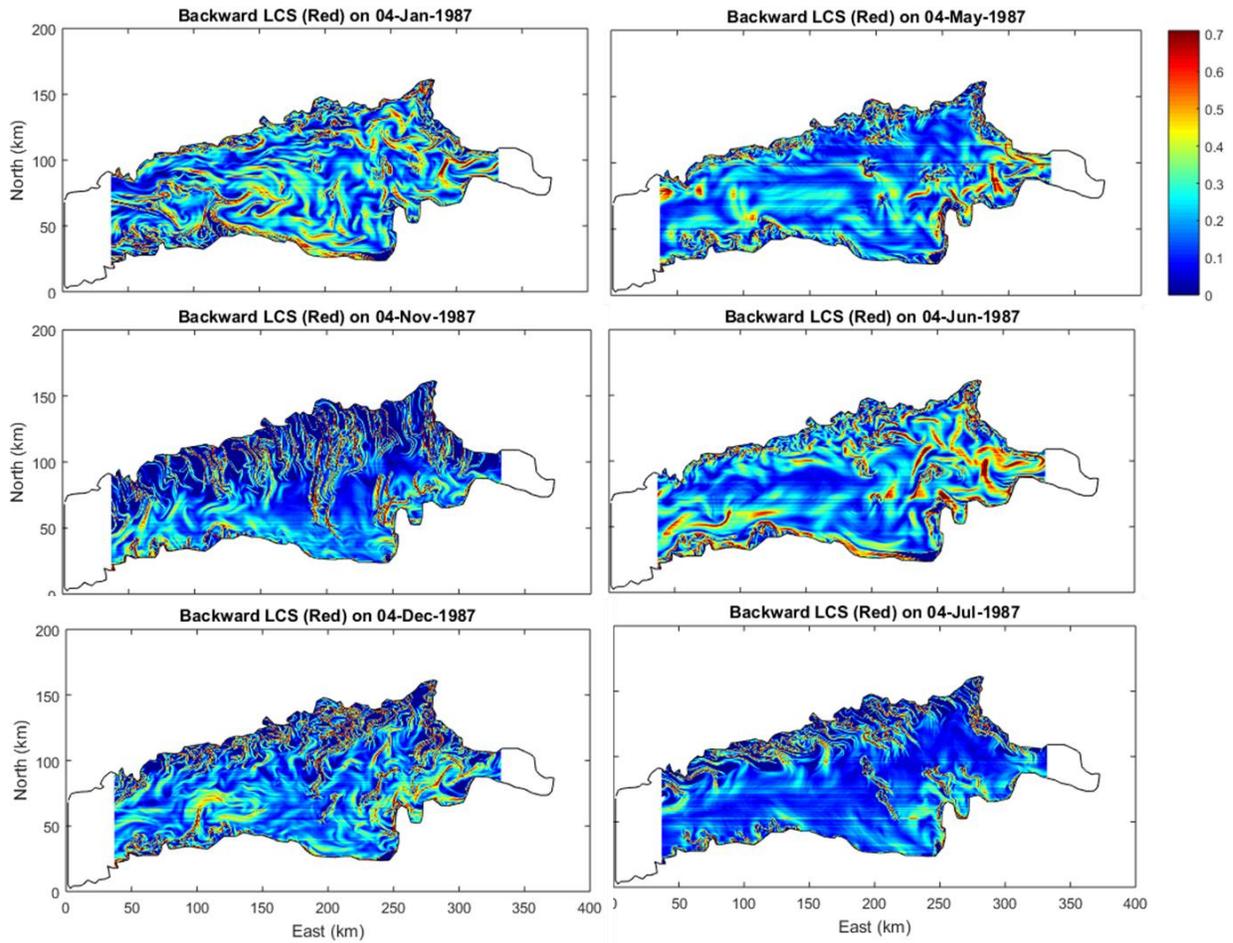

**Figure 1**. Normalised backward FTLE calculated for the entire Gulf of Finland (integration time is 3 days) during the windy (November, December and January, left column) and calm season (May, June and July, right column) of 1991.

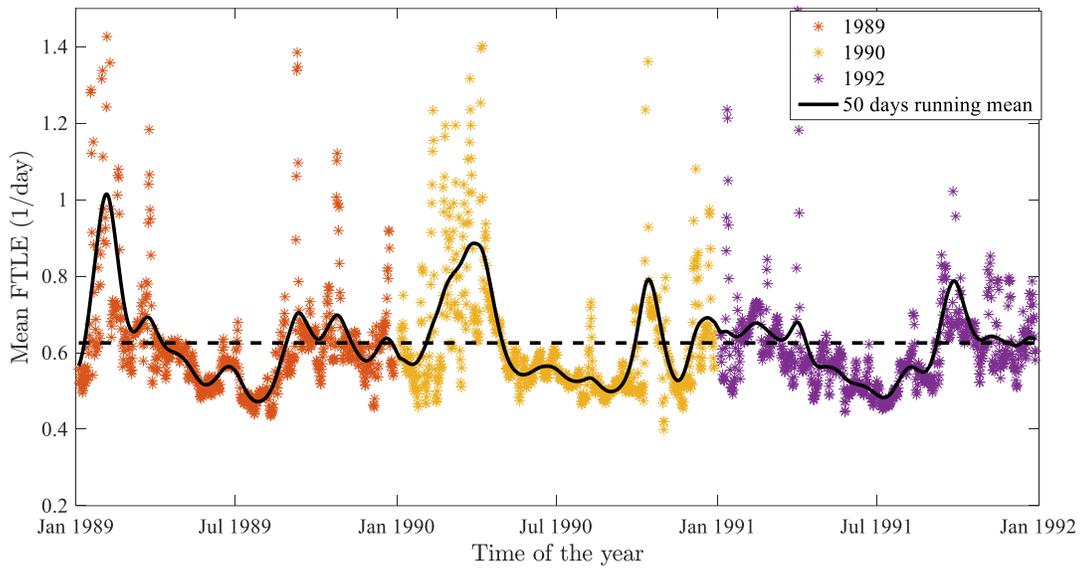

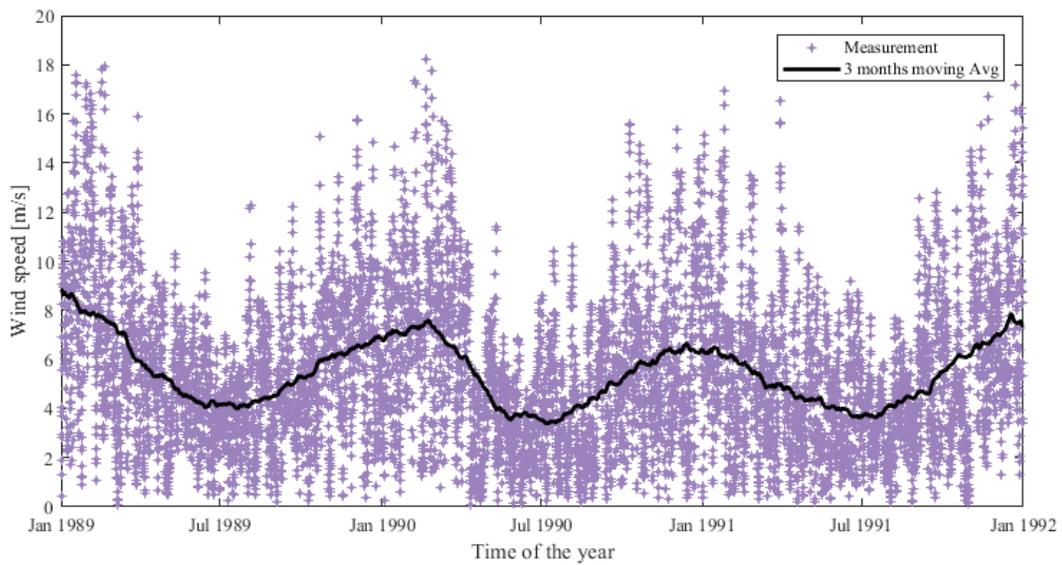

**Figure 2**. Gulf-wide spatial average of backward FTLE features, over the timespan 1989–1992, showing an increase in the mixing strength of the system during the windy season, and a decrease during the calm one. The dashed line represents the average over three years of the mean FTLE (upper panel). Hourly (thin blue line) and moving average (over a 3 months span) offshore wind speed at Kalbådagrund, located at 59.59°N, 25.36°E (lower panel).

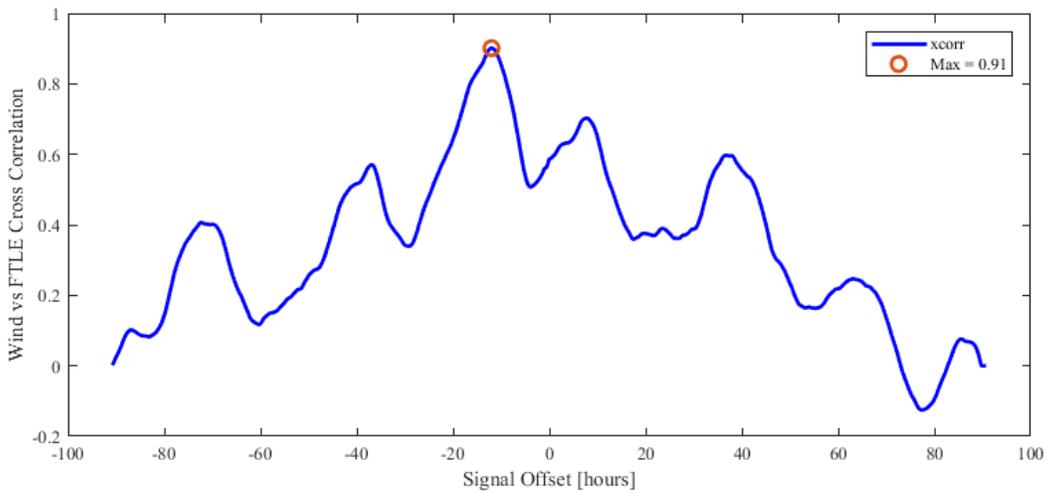

**Figure 3**. Cross-correlation of the mean FTLE and the wind speed at , Kalbådagrund in 1989–1992 as a function of their offset. The maximum correlation is 0.91, when the two signals are offset by about 20 h.

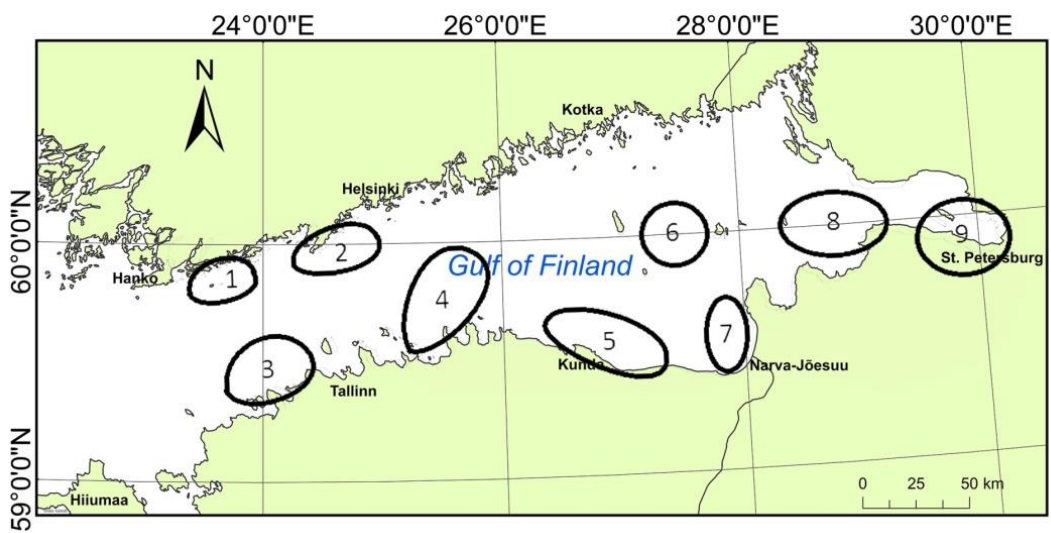

**Figure 4**. The 8 high-FTC areas as described in (Giudici and Soomere, 2013), which were considered when calculating localised average values of the backward FTLE features. These areas were shown to have high-clustering potential throughout the year by means of the FTC measure, and are thought to be hotspots for up/down-well mixing.

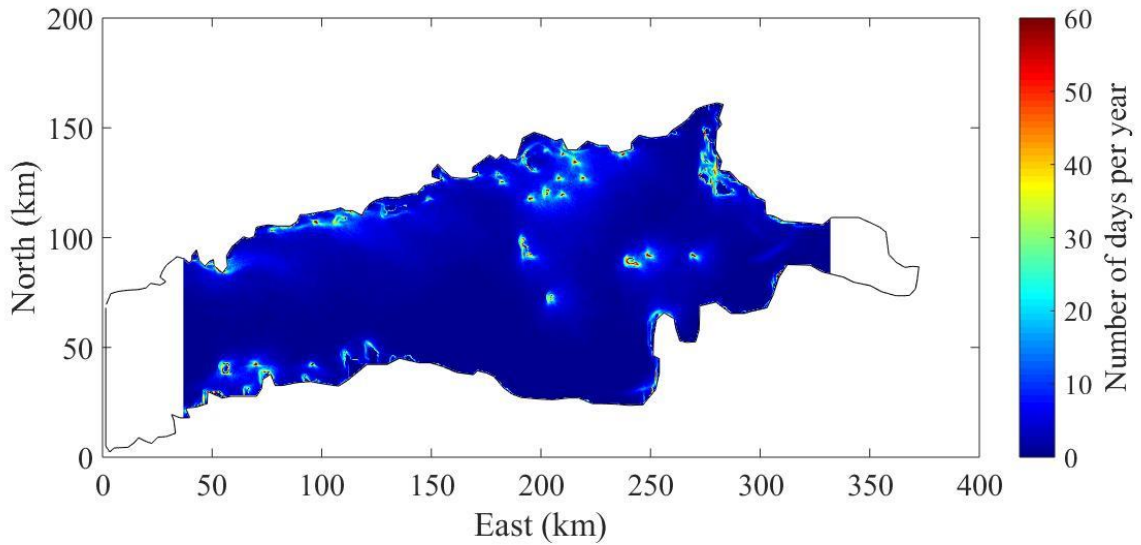

**Figure 5**. Locations where backward FTLE values are consistently high, during calendar years 1987–1991.

**Acknowledgements**

The project is supported from Estonian Research Council grants IUT33-3 and PRG1129, the Estonian Research Infrastructures Roadmap object Infotechnological Mobility Observatory (IMO) and through Australia Research Council Discovery Project grant DP190103379.

**References**

Alenius, P., Nekrasov, A., Myrberg, K., 2003. The baroclinic Rossby-radius in the Gulf of Finland. Cont. Shelf Res., 23(6), 563–573, https://doi.org/10.1016/S0278-4343(03)00004-9.

Andrejev, O., Sokolov, A., 1989. Numerical modelling of the water dynamics and passive pollutant transport in the Neva inlet. Meteorol. Hydrol. 12, 75–85 (in Russian).

Andrejev, O., Sokolov, A., 1990. 3D baroclinic hydrodynamic model and its applications to Skagerrak circulation modelling. In: Proc. 17th Conference of the Baltic Oceanographers. Norrköping, Sweden, pp. 38–46.

Andrejev, O., Myrberg, K., Alenius, P., Lundberg, P.A., 2004. Mean circulation and water exchange in the Gulf of Finland – a study based on three-dimensional modelling. Boreal Environ. Res., 9, 1–16.

Andrejev, O., Soomere, T., Sokolov, A., Myrberg, K., 2011. The role of spatial resolution of a three-dimensional hydrodynamic model for marine transport risk assessment. Oceanologia, 53(1-TI), 309–334, https://doi.org/10.5697/oc.53-1-TI.309.

Arrieta, J., Barreira, A., Tuval, I., 2017. Microscale patches of nonmotile phytoplankton. Phys. Rev. Lett. 114(12), Art. No. 128102, https://doi.org/10.1103/PhysRevLett.114.128102.

Balk, A.M., Falkovich, G., Stepanov, G., 2004. Growth of density inhomogeneities in a flow of wave turbulence, Phys. Rev. Lett. 92(24), Art. No. 244504, https://doi.org/10.1103/PhysRevLett.92.244504.

Bergström, S., Carlsson, B., 1994. River runoff to the Baltic Sea: 1950–1990. Ambio 23, 280–287.

Brentnall, S.J., Richards, K.J., Brindley, J., Murphy, E., 2003. Plankton patchness and its effect on larger-scale productivity. J. Plankton Res. 25, 121–140, https://doi.org/10.1093/plankt/25.2.121.

Chatterjee, R., Joshi, A.A., Perlekar, P., 2016. Front structure and dynamics in dense colonies of motile bacteria: Role of active turbulence. Phys. Rev. E 94(2), Art. No. 022406, https://doi.org/10.1103/PhysRevE.94.02240.

Dauch, T.F., Rapp, T., Chaussonnet, G., Braun, S., Keller, M.C., Kaden, J., Koch, R., Dachsbacher, C., Bauer, H.J., 2018. Highly efficient computation of Finite-Time Lyapunov Exponents (FTLE) on GPUs based on three-dimensional SPH datasets. Computers & Fluids, 175, 129–141, https://doi.org/10.1016/j.compfluid.2018.07.015.

Delpeche-Ellmann, N., Mingelaitė, T., Soomere, T., 2017. Examining Lagrangian surface transport during a coastal upwelling in the Gulf of Finland, Baltic Sea. J. Mar. Syst. 171, 21–30, https://doi.org/10.1016/j.jmarsys.2016.10.007.

Denissenko, P., Falkovich, G., Lukashuck, S., 2006. How waves affect the distribution of particles that float on a liquid surface, Phys. Rev. Lett. 97(24), Art. No. 244501, https://doi.org/10.1103/PhysRevLett.97.244501.

Elken, J., Raudsepp, U., Laanemets, J., Passenko, J., Maljutenko, I., Pärn, O., Keevallik, S., 2014. Increased frequency of wintertime stratification collapse events in the Gulf of Finland since the 1990s. J. Mar. Syst. 129, 47–55, https://doi.org/10.1016/j.jmarsys.2013.04.015.

Farazmand, M., Haller, G., 2012. Computing Lagrangian coherent structures from variational LCS theory. Chaos 22, Art. No. 01328, https://doi.org/10.1063/1.3690153.

Fennel, W., Radtke, H., Schmidt, M., Neumann, T., 2010. Transient upwelling in the central Baltic Sea. Cont. Shelf Res. 30, 2015–2026, https://doi.org/10.1016/j.csr.2010.10.002.

Froyland, G., Stuart, R.M., 2014. How well-connected is the surface of the global ocean? Chaos 24, Art. No. 033126, https://doi.org/10.1063/1.4892530.

Giudici, A., Kalda, J., Soomere, T., 2012. On the compressibility of surface currents in the Gulf of Finland, the Baltic Sea. In: Proceedings of the IEEE/OES Baltic 2012 International Symposium "Ocean: Past, Present and Future. Climate Change Research, Ocean Observation & Advanced Technologies for Regional Sustainability" (May 8–11, Klaipėda, Lithuania). IEEE Conference Publications, pp. 1–8, https://doi.org/10.1109/BALTIC.2012.6249178

Giudici, A., Soomere, T., 2013. Identification of areas of frequent patch formation from velocity fields. J. Coast. Res. Special Issue 65, 231–236, https://doi.org/10.2112/SI65-040.1.

Giudici, A., Soomere, T., 2014. Finite-time compressibility as an agent of frequent spontaneous patch formation in the surface layer: a case study for the Gulf of Finland, the Baltic Sea. Mar. Pollut. Bull. 89(1–2), 239–249, https://doi.org/10.1016/j.marpolbul.2014.09.053.

Giudici, A., Kalda, J., Soomere, T., 2018. Joint impact of currents and winds on the patch formation near the coasts of the Gulf of Finland. J. Coast. Res. Special Issue 85, 1156–1160, https://doi.org/10.2112/SI85-232.1.

Giudici, A., Kalda, J., Soomere, T., 2019. Generation of large pollution patches via collisions of sticky floating parcels driven by wind and surface currents. Mar. Pol. Bull. 141, 573–585, https://doi.org/10.1016/j.marpolbul.2019.02.039.

Granskog, A., Kaartokallio, H., Thomas, D.N., Kuosa, H., 2005. Influence of freshwater inflow on the inorganic nutrient and dissolved organic matter within coastal sea ice and underlying waters in the Gulf of Finland (Baltic Sea). Estuar. Coast. Shelf Sci. 65(1–2), 109–122, https://doi.org/10.1016/j.ecss.2005.05.011.

Haapala J., 1994. Upwelling and its influence on nutrient concentration in the coastal area of the Hanko Peninsula, entrance of the Gulf of Finland. Estuar. Coast. Shelf Sci. 38(5), 507–521, https://doi.org/10.1006/ecss.1994.1035.

Hadjighasem, A., Farazmand, M., Blazevski, D., Froyland, G., Haller, G., 2017. A critical comparison of Lagrangian methods for coherent structure detection. Chaos 27(5), Art. No. 053104, https://doi.org/10.1063/1.4982720.

Haller, G., 2001. Distinguished material surfaces and coherent structures in three-dimensional fluid flows. Physica D 149(4), 248–277, https://doi.org/10.1016/S0167-2789(00)00199-8.

Haller, G., 2002. Lagrangian coherent structures from approximate velocity data. Phys. Fluids 14(6), 1851–1861, https://doi.org/10.1063/1.1477449.

Haller, G., 2011. A variational theory of hyperbolic Lagrangian coherent structures. Physica D 240, 574–598, https://doi.org/10.1016/j.physd.2010.11.010.

Haller, G., 2015. Lagrangian coherent structures. Annu. Rev. Fluid Mech. 47, 137–162, https://doi.org/10.1146/annurev-fluid-010313-141322.

Hela I., 1976. Vertical velocity of the upwelling in the sea. Commentat. Phys.-Math., Soc. Scient. Fennica 46(1), 9–24, Helsinki.

Hernández-Carrasco, I., Orfila, A., Rossi, V. and Garçon V., 2018. Effect of small scale transport processes on phytoplankton distribution in coastal seas. Sci. Repts. 8, Art. No. 8613, https://doi.org/10.1038/s41598-018-26857-9.

Höglund, A., Meier, H.E.M., Broman, B., Kriezi, E., 2009. Validation and correction of regionalised ERA-40 wind fields over the Baltic Sea using the Rossby Centre Atmosphere Model RCA3.0. Technical Report No. 97, Rapport Oceanografi, 29 pp.

Huntley, H.S., Lipphardt, B.L., Jacobs, G., Kirwan jr., A.D., 2015. Clusters, deformation, and dilation: Diagnostics for material accumulation regions. J. Geophys. Res.-Oceans 120, 6622–6636, https://doi.org/10.1002/2015JC011036.

Huhn, F., von Kameke, A., Allen-Perkins, S., Montero, P., Venancio, A., Pérez-Muñuzuri, V., 2012. Horizontal Lagrangian transport in a tidal-driven estuary—Transport barriers attached to prominent coastal boundaries. Cont. Shelf Res., 39–40, 1–13, https://doi.org/10.1016/j.csr.2012.03.005.

Jacobs, G.A., Huntley, H.S., Kirwan jr., A.D., Lipphardt jr., B.L., Campbell, T., Smith, T., Edwards, K.,. Bartels, B., 2016. Ocean processes underlying surface clustering. J. Geophys. Res.-Oceans 121, 180–197, https://doi.org/10.1002/2015JC011140.

Kalda, J., Soomere, T., Giudici, A., 2014. On the finite-time compressibility of surface currents in the Gulf of Finland, the Baltic Sea. J. Mar. Syst. 129, 56–65, https://doi.org/10.1016/j.jmarsys.2012.08.010.

Katsanoulis, S., Haller, G., 2019. BarrierTool Manual. Retrieved from https://github.com/LCSETH.

Kononen, K., Nõmmann, S., 1992. Spatio-temporal dynamics of the cyanobacterial blooms in the Gulf of Finland, Baltic Sea. In: Carpenter, E.J., Capone, D.G., Rueter, J.G. (Eds.), Marine Pelagic Cyanobacteria: Trichodesmium and other Diazotrophs. NATO ASI Series (Series C: Mathematical and Physical Sciences), vol. 362. Springer, Dordrecht, pp. 95–113, https://doi.org/10.1007/978-94-015-7977-3_7.

Myrberg, K., Andrejev, O., Lehmann, A., 2010. Dynamic features of successive upwelling events in the Baltic Sea – a numerical case study. Oceanologia 52(1), 77–99, https://doi.org/10.5697/oc.52-1.077.

Lee, D.K., Niiler, N., 2010. Influence of warm SST anomalies formed in the eastern Pacific subduction zone on recent El Nino events. J. Mar. Res. 68, 459–477, https://doi.org/10.1007/s00382-015-2630-1.

Lehmann, A., Myrberg, K., Höflich, K. 2012. A statistical approach to coastal upwelling in the Baltic Sea based on the analysis of satellite data for 1990–2009. Oceanologia 54(3), 369–393, https://doi.org/10.5697/oc.54-3.369.

Leppäranta, M., Myrberg, K., 2009. Physical oceanography of the Baltic Sea. Springer, Chichester, UK, 378 pp.

Lips, I., Lips, U., Liblik, T., 2009. Consequences of coastal upwelling events on physical and chemical patterns in the central Gulf of Finland (Baltic Sea). Cont. Shelf Res., 29(15), 1836–1847, https://doi.org/10.1016/j.csr.2009.06.010.

Maljutenko, I., Raudsepp, U. 2019. Long-term mean, interannual and seasonal circulation in the Gulf of Finland - The wide salt wedge estuary or gulf type ROFI. J. Mar. Syst. 195, 1–19, https://doi.org/10.1016/j.jmarsys.2019.03.004.

Martin, A.P., 2003. Phytoplankton patchiness: the role of lateral stirring and mixing. Progr. Oceanogr. 57(2), 125–174, https://doi.org/10.1016/S0079-6611(03)00085-5.

Meier, H.E.M. 2007. Modeling the pathways and ages of inflowing salt- and freshwater in the Baltic Sea. Est. Coast. Shelf Sci. 74, 610–627, https://doi.org/10.1016/j.ecss.2007.05.019.

Meier, H.E.M., Döscher, R., Faxén, T., 2003. A multiprocessor coupled ice-ocean model for the Baltic Sea: application to salt inflow. J. Geophys. Res.-Oceans 108(C8), 32–73, https://doi.org/10.1029/2000JC000521.

Messie, M., Chavez, F.P., 2017. Nutrient supply, surface currents, and plankton dynamics predict zooplankton hotspots in coastal upwelling systems. Geophys. Res. Lett. 44(17), 8979–8986, Art. No. L074322, https://doi.org/10.1002/2017GL074322.

Murawski, J., Woge Nielsen, J., 2013. Applications of an oil drift and fate model for fairway design. In: Soomere, T., Quak, E. (Eds.), Preventive Methods for Coastal Protection: Towards the Use of Ocean Dynamics for Pollution Control. Springer, Cham Heidelberg, pp. 367–415, https://doi.org/10.1007/978-3-319-00440-2_11.

Olascoaga, M.J., Rypina, I.I., Brown, M.G., Beron-Vera, F.J., Koçak, H., Brand, L.E., Halliwell, G.R., Shay, L.K., 2006. Persistent transport barrier on the West Florida Shelf. Geophys. Res. Lett. 33(22), Art. No. L22603, https://doi.org/10.1029/2006gl027800.

Olascoaga, M.J., 2010. Isolation on the West Florida Shelf with implications for red tides and pollutant dispersal in the Gulf of Mexico. Nonlinear Process Geophys, 17(6), 685–696, https://doi.org/10.5194/npg-17-685-2010.

Olascoaga, M.J., Haller, G., 2012. Forecasting sudden changes in environmental pollution patterns. Proceedings of the National Academy of Sciences of the United States of America, 109(13), 4738–4743, https://doi.org/10.1073/pnas.1118574109.

D'Ovidio, F., Fernández.V., Hernández-García, E. López, C., 2004. Mixing structures in the Mediterranean Sea from finite-size Lyapunov exponents. Geophys. Res. Lett. 31(17), Art. No. L17203, https://doi.org/10.1029/2004GL020328.

Pavelson, J., Laanemets, J., Kononen, K., Nõmmann, S., 1997. Quasi-permanent density front at the entrance to the Gulf of Finland: Response to wind forcing. Cont. Shelf Res., 17(3), 253–265, https://doi.org/10.1016/S0278-4343(96)00028-3.

Peikert, R., Pobitzer, A., Sadlo, F., Schindler, B., 2014. A comparison of finite-time and finite-size Lyapunov exponents. In: Bremer, P.-T., Hotz, I.,Pascucci, V., Peikert, R. (Eds.), Topological Methods in Data Analysis and Visualization III: Theory, Algorithms, and Applications. Springer, Cham, pp. 187–200, https://doi.org/10.1007/978-3-319-04099-8_12.

Pichel, W.G., Churnside, J.H., Veenstra, T.S., Foley, D.G., Friedman, K.S., Brainard, R.E., Nicoll, J.B., Zheng, Q., Clemente-Colon, P., 2007. Marine debris collects within the North Pacific Subtropical convergence zone. Mar. Pollut. Bull. 54, 1207–1211, https://doi.org/10.1016/j.marpolbul.2007.04.010.

Rockwood, M.P., Loiselle, T., Green, M.A., 2019. Practical concerns of implementing a finite-time Lyapunov exponent analysis with under-resolved data. Exp. Fluids 60(4), Art. No. 74, https://doi.org/10.1007/s00348-018-2658-1.

Samuelsen, A., Hjøllo, S.S., Johannessen, J.A., Patel, R., 2012. Particle aggregation at the edges of anticyclonic eddies and implications for distribution of biomass. Ocean Sci. 8, 389–400, https://doi.org/10.5194/os-8-389-2012.

Samuelsson, P., Jones, C.G., Willén, U., Ullerstig, A., Gollvik, S., Hansson, U., Jansson, C., Kjellström, E., Nikulin, G., Wyser, K., 2011. The Rossby Centre Regional Climate Model RCA3: Model description and performance. Tellus 63A 4–23, https://doi.org/10.1111/j.1600-0870.2010.00478.x.

Shadden, S.C., 2011. Lagrangian coherent structures. In Grigoriev, R. (Ed.), Transport and Mixing in Laminar Flows. Wiley, pp. 59–89, https://doi.org/10.1002/9783527639748.ch3.

Savijärvi, H., Niemela, S., Tisler, P., 2005, Coastal winds and low-level jets: simulations for sea gulfs. Q. J. Roy. Meteor. Soc. B, 131(606), 625–637, https://doi.org/10.1256/qj.03.177.

Soomere, T., 2013. Applications of the inverse problem of pollution propagation. In Soomere, T., Quak, E. (Eds.), Preventive Methods for Coastal Protection: Towards the Use of Ocean Dynamics for Pollution Control. Springer, pp. 319–366, https://doi.org/10.1007/978-3-319-00440-2_10.

Soomere, T., Keevallik, S., 2003. Directional and extreme wind properties in the Gulf of Finland. Proc. Estonian Acad. Sci. Eng., 9(2), 73–90.

Soomere, T., Myrberg, K., Leppäranta, M., Nekrasov, A., 2008. The progress in knowledge of physical oceanography of the Gulf of Finland: a review for 1997–2007. Oceanologia 50(3), 287–362, 2008.

Soomere, T., Delpeche, N., Viikmäe, B., Quak, E., Meier, H.E.M., Döös, K., 2011. Patterns of current-induced transport in the surface layer of the Gulf of Finland. Boreal Environ. Res. 16 (Suppl. A), 49–63.

Soomere, T., Döös, K., Lehmann, A., Meier, H.E.M., Murawski, J., Myrberg, K., Stanev, E., 2014. The potential of current- and wind-driven transport for environmental management of the Baltic Sea. Ambio 43, 94–104, https://doi.org/10.1007/s13280-013-0480-9.

Suara, K., Chanson, H., Borgas, M., Brown, R.J., 2017. Relative dispersion of clustered drifters in a small micro-tidal estuary. Estuar. Coast. Shelf Sci., 194, 1–15, https://doi.org/10.1016/j.ecss.2017.05.001.

Suara, K., Brown, R., Chanson, H., 2019. Characteristics of flow fluctuations in a tide-dominated estuary: Application of triple decomposition technique. Estuar. Coast. Shelf Sci. 218, 119–130, https://doi.org/10.1016/j.ecss.2018.12.006.

Suara, K., Khanarmuei, M., Ghosh, A., Yu, Y., Zhang, H., Soomere, T., Brown, R.J. 2020. Material and debris transport patterns in Moreton Bay, Australia: The influence of Lagrangian coherent structures, Sci. Total Environ. 721, Art. No. 137715, https://doi.org/10.1016/j.scitotenv.2020.137715.

Tang, W., Chan, P.W., Haller, G., 2011. Lagrangian coherent structure analysis of terminal winds detected by LIDAR. Part II: structure, evolution and comparison with flight data. J. Appl. Meteorol. Climatol. 50, 325–338, https://doi.org/10.1175/2011JAMC2689.1.

Vandenbulcke, L., Beckers, J.-M., Lenartz, F., Barth, A., Poulain, P.-M., Aidonidis, M., Meyrat, J., Ardhuin, F., Tonani, M., Fratianni, C., Torrisi, L., Pallela, D., Chiggiato, J., Tudor, M., Book, J.W., Martin, P., Peggion, G., Rixen, M., 2009. Super-ensemble techniques: Application to surface drift prediction. Progr. Oceanogr. 82, 149–167, https://doi.org/10.1016/j.pocean.2009.06.002.

Viikmäe, B., Soomere, T., Viidebaum, M., Berezovski, A., 2010. Temporal scales for transport patterns in the Gulf of Finland. Estonian J. Eng. 16(3), 211–227.

Vucelja, M., Falkovich, G., Fouxon, I., 2007. Clustering of matter in waves and currents, Phys. Rev. E 75(6), Art. No. 065301, https://doi.org/10.1103/PhysRevE.75.065301.

Zhurbas, V., Väli, G., Kuzmina, N., 2019. Rotation of floating particles in submesoscale cyclonic and anticyclonic eddies: a model study for the southeastern Baltic Sea. Ocean Sci. 15(6), 1691–1705, https://doi.org/10.5194/os-15-1691-2019.

Wasmund, N., Nausch, G., Voss, M., 2012. Upwelling events may cause cyanobacteria blooms in the Baltic Sea. J. Mar. Syst. 90(1), 67–76, https://doi.org/10.1016/j.jmarsys.2011.09.001.